# Knowledge Management

Mohsen Gerami

The Faculty of Applied Science of Post and Communications
Danesh Blv, Jenah Ave, Azadi Sqr, Tehran, Iran.
Postal code: 1391637111

*Abstract-This paper discusses the important process of knowledge and its management, and differences between tacit and explicit knowledge and understanding the culture as a key issue for the successful implementation of knowledge management, in addition to, this paper is concerned with the four-stage model for the evolution of information technology (IT) support for knowledge management in law firms.*

## I. INTRODUCTION

The move from an industrially-based economy to knowledge or information-based one in the 21st Century demands a top-notch knowledge management system to secure a competitive edge and a capacity for learning. Currently, governments around the world, multinational corporations, and a multitude of companies are interested, even concerned with the concept of knowledge management. The new source of wealth is knowledge, and not labor, land, or financial capital. It is the intangible, intellectual assets that must be managed. The key challenge of the knowledge-based economy is to foster innovation [1].

From a management perspective the key difference between information and knowledge is that information is much more easily identified, organized and distributed. Knowledge, on the other hand, cannot really be managed because it resides in one's mind. Thus, KM is essentially limited to creating the right conditions for individuals to learn (using information and experiencing the world) and apply their knowledge to the benefit of the organization [2]

### WHAT IS KNOWLEDGE AND HOW DO WE USE IT

*"Knowledge is power."*

*"But mere knowledge is not power; it is only possibility. Action is power; and its highest manifestation is when it is directed by knowledge." (Francis Bacon)*

Knowledge is:

DATA:                            Facts

INFORMATION:        Data organised for a purpose that which reduces uncertainty

KNOWLEDGE:            That which enlightens decisions and action [3]

Or we can write the knowledge process as:

Data:                          Flight AZ240. Arrival time 7:40 a.m

Information:              I'm booked on AZ240 & it's 60 minutes late

Understanding:  It might make up the time but it probably won't

Relevance:              It might make up the time but it probably won't

Knowledge:        It happened before & we met @ the airport or I may be able to fly BA instead

Utilisation:        I'll call John and ask him to come or tell him my new flight arrangements

Knowledge can only be volunteered, not conscripted. We always know more than we can tell, and tell more than we can write down. We only know what we know when we need to know it (David Snowden) [4]
In others words we have to understand levels of knowledge

[Symbols (+syntax) ·

Data (+meaning) ·

Information (+context, experiences) ·

Knowledge (+applied) ·

Know- how (+will to do) ·

Action (+adapted implementation to specific context) ·

Competence (+unique combination) ·

Competitiveness]

to improve operational and strategic knowledge management [5]

## II.  WHAT IS KNOWLEDGE MANAGEMENT

Knowledge management is a systematic process for acquiring, organizing, sustaining, applying, sharing, and renewing both tacit and explicit knowledge to enhance the organizational performance, increase organizational adaptability, increase values of existing products and services, and/or create new knowledge-intensive products, processes and services                    [6]

Knowledge Management is the process of developing knowledge and accumulating it in the organisational capital wherever possible. Knowledge Management is helping all managers to establish knowledge resource management as part of        their        toolkit        [3]

Knowledge management is the process of making relevant information available quickly and easily for people to use productively. For KM to move from ideas to implementation, the definition of KM needs to address:

- Creating, sharing, and reusing knowledge

- Understanding the relevance of different information as determined by the customer

- Training for KM methods and services

- Incorporating cultural aspects of KM into operations

- Responding to funding and chargeback issues

The Knowledge Management Process (a.k.a. "Doing Work")

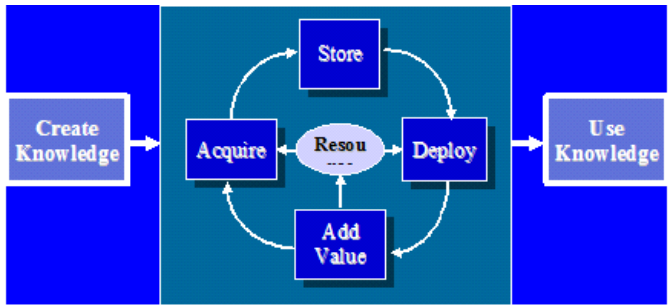

Figure 1. Knowledge Management Process

Knowledge management is the process of making relevant information available quickly and easily for people to use productively [7].

The value of knowledge is derived from the value of the decisions with which it is associated. The measurement of KM success is therefore related to improved decision making and the achievement of objectives. Some measures may be objective; others may be more subjective such as attitude surveys among stakeholders. Knowledge Management is making better decisions by understanding the knowledge ingredients for decision making.

In other words:

- On a personal level, knowledge can be what we want it to be. For an organisation, it is what we need for decisions and actions.

- Knowledge is found in people, processes and information, where information includes images and all forms of multi-media

- To understand the development of knowledge, the knowledge spiral is a very effective starting point

- Management means thinking of knowledge as a resource. Some resources may justify the description as Intellectual Capital.

- The effective implementation of knowledge management is directly related to Change Management [3]

## III.  TWO KINDS OF KNOWLEDGE

Knowledge is intangible, dynamic, and difficult to measure, but without it no organization can survive.

- Tacit: or unarticulated knowledge is more personal, experiential, context specific, and hard to formalize; is difficult to communicate or share with others; and is generally in the heads of individuals and teams.

- Explicit: explicit knowledge can easily be written down and        codified        [1]

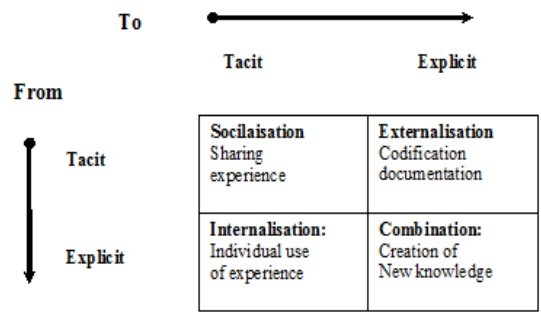

Figure 2. Knowledge process

We can transfer tacit knowledge through mechanisms of socialization, mentor ships, apprenticeships, face-to-face communication. Since knowledge may be an organization's only sustainable competitive advantage, it is very important to capture tacit knowledge. Intranets and e-mail help knowledge flow through an organization. Tacit knowledge often moves laterally through informal channels of communication (communities of practice). For example, those groups that hang around the coffee pot or the coffee machine -- they are exchanging knowledge, just as the smokers huddled near the entrance to the building at break time. The information that is passed in this way is very important because it is useful for helping people to get their work done more effectively, in part, because nobody is willing to question or think about it very much. Communities of practice must have their place in a comprehensive knowledge management effort [1].

Different cultural approaches different KM, Europe concentrates on knowledge *valuation*; United States on the *management* of explicit knowledge; Both result from the Cartesian perspective of a differentiation between mind and body, subject and object, knowledge and the knowing person. The Japanese approach concentrates on the *creation of knowledge,…* in a perspective of unity of mind and body.

Occident puts the emphasis on *explicit* knowledge, Japan on *implicit* one                                          [5]

### CULTURE - AND CULTURE CHANGE

Understanding the culture is a key issue for the successful implementation of knowledge management. The existing culture can amplify or inhibit knowledge management projects.

Knowledge management initiatives can support a change in organisational culture                       [3] Changing the culture is imperative. To create a climate in which employees volunteer their creativity and expertise, managers need to look beyond the traditional tools at their disposal: finding ways to build trust and develop fair process. That means getting the gatekeepers to facilitate the flow of information rather than hoard it and offering rewards and incentives.

The elements of fair process are simple: 1. Engage people's input in decisions that directly affect them. 2. Explain why decisions are made the way they are. 3. Make clear what will be expected of employees after the changes are made. Fair process may sound like a soft issue, but it is crucial to building trust and unlocking ideas.

Adds Buckman Laboratories' Koskiniemi: "Successful knowledge sharing is 90 percent cultural, 5 percent tools and 5 percent magic. All the technology and tools in the world won't make you a knowledge-based organization if you do not establish a culture that believes in sharing."

Organizations must offer a high level of psychological safety and capacity for openness.

Rewards and incentives signal what behaviors and outcomes are most valued by management. It should not be surprising that knowledge accumulation and sharing are not valued. Management sends strong signals through its compensation policies; different roles are perceived of value according to their allocated compensation. So be careful sending mixed signals. But culture is more than just compensation, and it is responsive to influences other than paychecks. Management sends signals about what is important through its recruiting priorities, promotions, and, possibly more than anything, through its own behavior. These deeply embedded cultural assumptions are significant [1]

### KM AND LAW

Here will define a four-stage model for the evolution of information technology (IT) support for knowledge management. The purpose of the model is both to understand the current situation in a firm in terms of a specific stage and to develop strategies for moving to a higher stage in the future. The model is applied to law firms where the knowledge of professional experts is a core asset and the careful management of this asset has special importance [8]

### Law Firms

A law firm can be understood as a social community specializing in speed and efficiency in the creation and transfer of legal knowledge [9]. Many law firms represent large corporate enterprises, organizations or entrepreneurs with a need for continuous and specialized legal services that can only be supplied by a team of lawyers. The client is a customer of the firm, rather than a particular lawyer. Relationships with clients tend to be enduring [10].

Lawyers can be defined as knowledge workers. They are professionals who have gained knowledge through formal education (explicit) and through learning on the job (tacit). After completing their advanced educational requirements, most professionals enter their careers as associates in law. In this role, they continue to learn and, thus, they gain significant tacit knowledge through 'learning by doing' [8] Lawyers work in law firms and law firms belong to the legal industry. The legal industry will change rapidly in the future because of three important trends. First, global companies increasingly seek out law firms that can provide consistent support at all business locations and integrated cross-border assistance for significant mergers and acquisitions as well as capital market transactions. Second, client loyalty is decreasing as companies increasingly base purchases of legal services on a more objective assessment of their value, defined as the benefits net of price. Finally, new competitors have entered the market, such as accounting firms and Internet-based legal services firms [11] Montana was not convinced that law firms will change, arguing that law stands out as an anachronism in the age of knowledge management. Law is entirely man-made: there are no hidden physical principles. A person researching some question of law ought to be able to derive an answer with certainty quickly and easily. According to Montana nothing is further from the truth [12]

### The Knowledge Management Technology Stage Model

The stages of knowledge management technology is a relative concept concerned with IT's ability to process information for knowledge work. IT at later stages is more useful to knowledge work than IT at earlier stages. The relative concept implies that IT is more directly involved in knowledge work at higher stages and that IT is able to support more advanced knowledge work at higher stages.

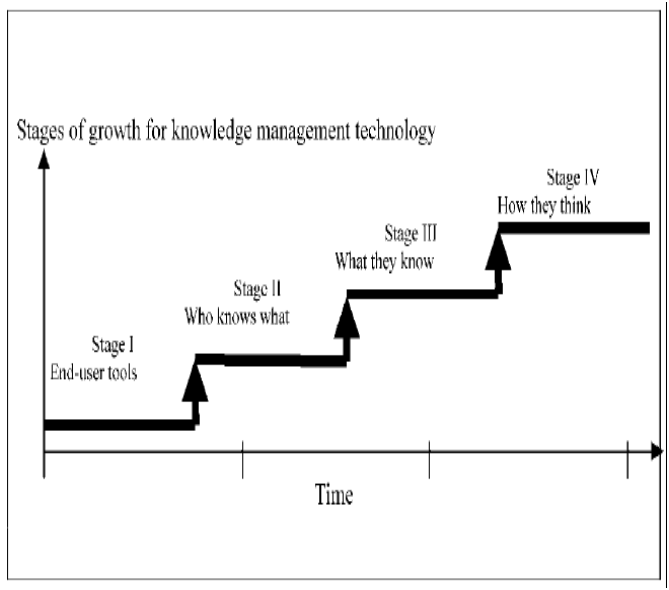

Figure3. The stages of the growth model for knowledge management technology

Stage I: end-user tools (people to technology)

End-user tools will be made available to lawyers. This means a capable networked PC on every desk or in every briefcase, with standardized personal productivity tools (word processing and presentation software) so that documents can be exchanged easily throughout a company. A widespread dissemination and use of end-user tools among knowledge workers in the company is to take place.

Stage II: who knows what (people to people)

Information about who knows what will be made available to lawyers. It aims to record and disclose who in the organization knows what by building knowledge directories. Often called Yellow Pages, the principal idea is to make sure knowledgeable people in the organization are accessible to others for advice, consultation or knowledge exchange. Knowledge-oriented directories are not so much repositories of knowledge-based information as gateways to knowledge.

Stage III: what they know (people to documents)

Information from lawyers will be stored and made available to colleagues. Here data-mining techniques will be applied to find relevant information and combine information in data warehouses. One approach is to store project reports, notes, recommendations and letters from each lawyer in the firm. Over time, this material will grow rapidly, making it necessary for a librarian or a chief knowledge officer to organize it.

Stage IV: what they think (people to systems)

An IS solving knowledge problems will be made available to lawyers. Artificial intelligence will be applied in these systems. For example, neural networks are statistically oriented tools that excel at using data for classifying cases into one category or another. Another example is expert systems that can enable the knowledge of one or a few experts to be used by a much broader group of lawyers who need the knowledge [8]

Knowledge management strategies focusing on personalization could be called communication strategies, because the main objective is to foster personal communication between people. Core IT systems with this strategy are Yellow Pages (directories of experts, who-knows-what systems and people-finder databases) that show inquirers who they should talk to regarding a given topic or problem. The main disadvantages of personalization strategies are a lack of standards and the high dependence on communication skills and the will of the professionals. Such disadvantages make firms want to advance to stage III. In stage III, independence in time among knowledge suppliers and knowledge users is achieved [13]

KM AND FUTURE SCENARIOS

One of the major problems with governments, corporations, companies, organizations, and private citizens is that they have no concept of the future and never think about ramifications? Future studies must be figured into an organization's overall knowledge management system because to sustain a commitment over the course of months and years, people need to have awareness of the whole and understand the direction an organization is going. The challenge of organizational strategy and purpose is to revitalize and rethink the organization's business focus, and figure out where it is heading. To expect ongoing knowledge creation, it must have some relevance to the future you are creating. Therefore, a future element must be ever-present. If you can only offer the wholesale version with precautions thrown in, it is better than the present reality. Future scenarios should not only be for the haves.

Knowledge Management must somehow be connected to future studies for at least one significant reason and that is because to have a knowledge management system, it presupposes the ongoing creation of new knowledge. The challenge of organizational strategy and purpose is to revitalize and rethink the organization's business focus, and figure out where it is heading. Peter Drucker's Theory of Business can also be brought into this analysis because he believed that there must be significant focus put on defining the environment, mission, and core competencies needed to accomplish that mission. "If the attitudes brought forth are genuinely heartfelt, if managers and especially top managers can increase their vulnerability by exposing their own deepest aspirations and assumptions, if people can feel part of a larger creative process shaping their industry and society, and if all this can be tied to people's commitment to creating a future about which they deeply care - then intellect and spirit align, and energy is not only released but focused." [1]

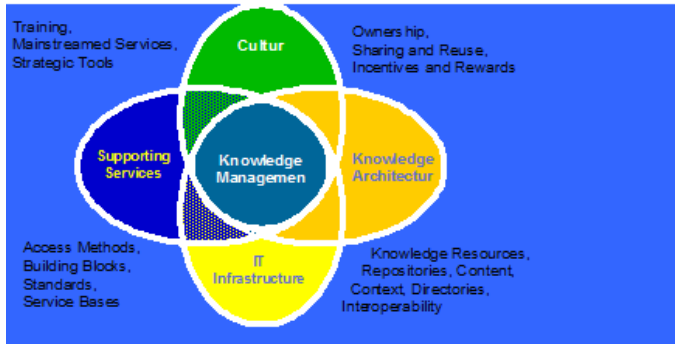

Figure 4. KM Success Factors

NASA Agency, 2000

## IV. CONCLUSION

Much of the confusion and disappointment concerning KM comes from a confusion between information and knowledge because knowledge is not linked to action, even by KM experts. There is no clarity. People are investing in systems to capture, organize, and disseminate information, and then calling it "knowledge." But knowledge cannot, by definition, be converted into an object and "given" from one person to another. Knowledge only diffuses when there are learning processes whereby human beings develop new capacities for effective action. Information technology, while critical for enabling the spread of information, cannot capture and store knowledge. Only people can do that [1].

Knowledge management must be seen as a priority which will enhance academic activity. Training for IT and information literacy is needed. Existing data sources must be managed well. Knowledge management must relate to personal and unit goals as well as institutional goals. Knowledge sharing must be fostered. Responsibility for the coordination of the whole of knowledge management is required. The process involves fundamental change which is evolutionary.

"We don't know one millionth of one percent of anything." Thomas Edison.